# Scented Dice: New interaction qualities for ideating connected devices

Albrecht Kurze

Chemnitz University of Technology, Albrecht.Kurze@informatik.tu-chemnitz.de

Much research has been done around creating multisensory ideation and prototyping tools. We relate our own multisensory tool, the *Loaded Dice*, to this domain. We briefly explain what sensing and actuating possibilities the Loaded Dice already have (including thermal sensations), and how they are methodically embedded in workshops using an extended interaction vocabulary to characterize and ideate multisensory experiences. We briefly ponder simple technical ways to add smell as an output modality, and discuss why and how smell capabilities will enrich ideation workshops in co-design.

## 1 INTRODUCTION

A couple of years ago we designed and developed the *Loaded Dice* [10,11], a multisensory hybrid toolkit to ideate IoT devices and scenarios in different contexts, e.g. for the 'smart' home, and with different groups of co-designers [3,4,9,10]. The Loaded Dice filled a gap between analog, non-functional tools, often card-based, e.g. *KnowCards* [1], and functional but tinkering based tools, e.g. *littleBits* [2] for multisensory exploration, ideation and prototyping. The Loaded Dice offer easy and ready to use inputs and outputs for different senses, e.g. light and sound as well as vibration, motion and temperature. The Loaded Dice also allows generating new connected sensor-actuator combinations very quickly in co-design workshops. Our workshops often brought up a number of unconventional ideas of multisensory interactions and usage scenarios - often far beyond 'ordinary' inputs and especially outputs. The "We(a)ther Bird" [8,10] and the "Inflatable Cat" [4] are to mention here as ideated poetic and extraordinary interaction experiences.

While we had the ability to register and create thermal sensations right from the initial design of our ideation tool, we have not yet covered taste or smell. Addressing new senses, at least smell, promises to enrich ideation for interactions with and communication through (connected) devices.

We hope in the workshop for inspiration and discussion
  a) of feasible, easily achievable technical implementations for augmenting an existing multisensory ideation tool, the Loaded Dice
  b) how new multisensory interaction qualities, i.e. smell, can stimulate exploration and scenario driven ideation for connected devices

## 2 BACKGROUND AND STATE OF THE ART MULTISENSORY IDEATION

### 2.1 Loaded Dice

The Loaded Dice are a set of two cubical devices, figure 1. One cube incorporates sensors, the other cube actuators suitable for multisensory and multimodal environmental and user interactions, figure 2. Each cube has six sides, offering six sensors or actuators in a cube, one on each side (multi-sided and multi-functional). Both cubes are wirelessly connected. One cube senses a specific sensor value, normalizes it meaningfully, transmits it and then the other cube actuates it mapped on an output. Every sensor-in and actuator-out combination is possible (6x6=36 in total). To engage spontaneous exploration and prototyping, the tool had to offer an easy way to re-combine sensors and actuators. The form of a cube communicates the intuitive reading that the top side is active - like a die. Using two cubes allows for fast one-to-one pairing between the active sides, to experience the effect.

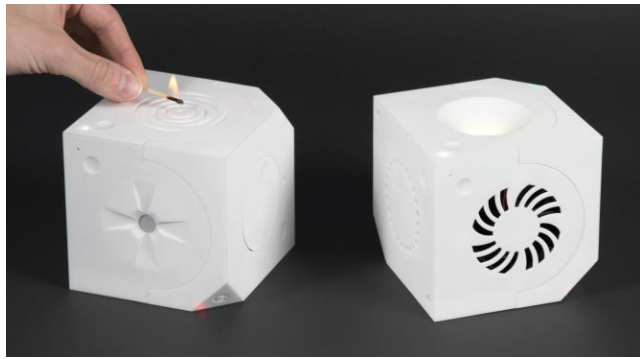

Figure 1: Loaded Dice, example of devices in use, turning heat into light (sensor die with temperature sensor active and actuator die with power LED active) [11]

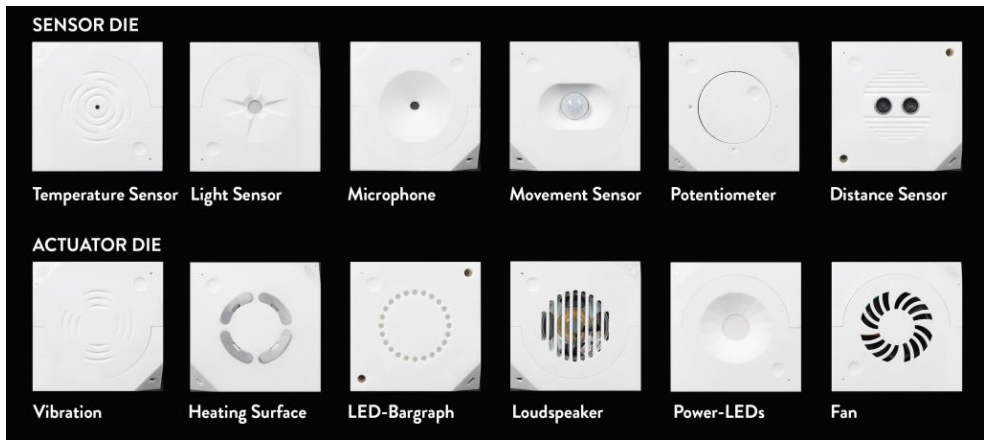

Figure 2: Faces and functions of the Loaded Dice – sensors and actuators [10]

## 2.2 Methodical embedding and interaction vocabulary

We use the Loaded Dice methodically embedded as Cards'n'Dice in combination with a card set to guide the process with co-designers in a workshop [4]. The card set also includes property cards that describe sensory qualities. Based on Diefenbach's interaction vocabulary [5] with pairs of adjectives we developed a slightly adapted and extended version with a few more pairs to match the IoT design space [4]. These cards relate to how local interaction with devices and through devices are conveyed and perceived by users. Examples are approximate–precise, casual–attention grabbing, direct–mediated, friendly–angry, graded–binary, incidental–targeted, instant–delayed, objective—poetic, powerful–gentle, private–public, slow–fast, soft–angry, etc.

## 3 SCENT AS INSPIRATION FOR DESIGN

Prior work, e.g. [7,12], has already researched how smell might be associated to non-verbal channels of communication and how it might characterized. E.g. [6,7] give hints that smell might also be associated to specific materiality or even shape of objects. This can be a good starting point for the smell of our cubical ideation devices. [12] researched the association of smell to a vocabulary of different overall qualities and even certain experienced emotions based on reported stories of participants. The covered overall qualities there were further differentiated in positive (pleasant, fresh, sweet, clean, mouthwatering) or negative (unpleasant, penetrating, heavy, foul, nauseous). The researchers differentiated the experienced emotions in the stories between positive, ambivalent and negative emotions:

- positive: happy, pleased, joyous, delighted, excited;
- ambivalent: happy, excited, enthusiastic, joyous, serene.
- negative: uncomfortable, disgusted, distressed, miserable, taken aback

Overall, the prior work shows the potential of smell for use in technology for non-verbal communication to the users. However, it is not yet clear, how specific interactions qualities might be mapped to specific olfactory sensations. E.g. [12] has already brought up the idea of a "reminder alert with smell" but without further characterization how such an interaction might be specified in detail, e.g. using an interaction vocabulary.

## 4 POTENTIAL AND CONSIDERATIONS FOR NEXT GENERATION MULTISENSORY IDEATION

In our workshops, we found some repeating themes when it comes to a mapping between certain interaction characteristics and suitable sensing as well as actuating possibilities. For example, the thermo-element was not only associated with warmth literally but also with 'love' in a poetic way. Participants often chose non-visible and non-audible modalities for 'private' interactions, covered and not easily perceivable by others, only noticeable to a mentioned one. Often we observed participants holding the actuator cube with the fan activated in front of their face, enjoying a cool breeze blowing directly in their face. However, it is just pure, odorless air.

It probably does not need much to add a simple **smell output** to the actuator cube of the Loaded Dice. Even the already included technical components might offer a basic mechanism for a controlled release of a single scent. The included thermoelectric Peltier element might heat up on demand an odor concentrate stored in the inside of the cube. The fan then blows out the scented air as gradually as needed – either directly to a participant or just in the room air when resting on a desk. In our experience, gained in numerous workshops, a sensory sensation does not need to be perfect, at least for ideation. It is about the idea and the core concept behind it. A demonstration of a technical possibility as a stimulus can be enough to trigger thinking about other possibilities and uses, including those that a simple ideation tool ultimately cannot provide.

## 5 MAPPING NEW SENSES TO THE INTERACTION VOCABULARY FOR IDEATION

Why is smell as an output modality interesting for ideation interaction and communication scenarios? Scents might probably cover a wide range of what can be described using the already existing interaction vocabulary. Smell might be even a perfect fit for some vocabulary entries. Using smell capabilities in ideation workshops might bring up preferred but until now undiscovered preferences of multisensory output for different use cases.

*Imaginable examples* of using smell and the according mapping to the interaction vocabulary to characterize such interactions are:

**slow or fast; instant or delayed:** While a scent might come up slowly, even delayed until reaching a certain concentration in the air, it also might come up very quickly, nearly instant. However, a 'smell-OFF' might be heavily delayed, it might take a while for a scent to slowly decay or thin out in a room, especially if not actively neutralized.

**constant (probably not inconstant):** A scent might be constant. A single smelly notification might be noticeable much longer than a short beep. You might smell the notification when you come back into a room, even if you have been away for minutes or even hours.

**graded, maybe binary; more fluent than stepwise; more approximate than precise; gentle or powerful; covered or apparent; incidental vs. targeted:** A scent might be very gentle, barely noticeable, or very powerful, sometimes even overwhelming. The incidental, casual character of smell is a possible special strength. Nosing a smell does not need a user's gaze or concentrated listening.

**direct and mediated:** A smell might hit you directly, bypassing other filters. It also might be mediated, soaked into a textile, in a cover etc. absorbing the smell and then mediating it.

**spatial proximate or separated:** Smell propagates through the air. A scent might take a few seconds to cross a room. The closer to the source of scent you are the quicker and more noticeable it will be. Airtight spatial separations (a door or box etc.) might block a scent.

**strange as well as familiar:** A specific scent can be very familiar, something you nose every day, or maybe every time in a specific context (e.g. the smell of a subway station or of rain) – or it can be something extraordinary – positive or negative, e.g. a chemical pungent smell indicating extreme danger.

**factual as well as metaphorical; probably more poetic than prosaic:** The smell of roasted and brewed coffee in the air might be more than just a notification that the coffee is ready and waits upstairs. It might also be associated with positive emotions, with a remembering of a nice afternoon with friends in this special French cafè on the corner or the smell of a weekend in grandma's countryside house. The brand of fragrance of a beloved one in the air might be more poetic and less disturbing than a "ding-dong"-sound, a blinking light or a buzzing of a vibration motor – as a notification or simply as a subtle "I'm thinking of you, darling."

## 6 CONCLUSION

Smell as an additional interaction quality might turn the Loaded Dice into *Scented Dice*, offering new and interesting possibilities for interactions in exploration, ideation, and prototyping co-design workshops.

## ACKNOWLEDGMENTS

This research is funded by the German Ministry of Education and Research (BMBF), grant FKZ 16SV7116.